\documentclass{aa}
\usepackage{graphicx}
\usepackage{amssymb}

\newcommand{\ebv}{E(B - V)}

\newcommand{\HI}{{\ion{H}{1}}}

\newcommand{\kms}{$\,$km$\,$s$^{-1}$}

\newcommand{\kmsMp}{km s$^{-1}$ Mpc$^{-1}$}
\newcommand{\mJybeam}{mJy beam$^{-1}$}
\newcommand{\msun}{{$M_\odot$}}

\newcommand{\tspin}{$T_{\rm spin}$}

\def\HI{\ion{H}{i}}

\def\OIII{\ion{O}{iii}}
\def\OII{\ion{O}{ii}}
\def\OI{\ion{O}{i}}
\def\NII{\ion{N}{ii}}
\def\SII{\ion{S}{ii}}
\def\FeVII{\ion{Fe}{vii}}
\def\FeX{\ion{Fe}{x}}
\def\CaV{\ion{Ca}{v}}
\def\emph#1{{\sl #1}}
\newcommand{\ltsima} {$\; \buildrel < \over \sim \;$}
\newcommand{\gtsima} {$\; \buildrel > \over \sim \;$}
\newcommand{\lta} {\lower.5ex\hbox{\ltsima}}
\newcommand{\gta} {\lower.5ex\hbox{\gtsima}}
\newcommand{\lala}{$\lambda\lambda$}

\newcommand{\hb}{H$\beta$}

\newcommand{\ha}{H$\alpha$}
\newcommand{\dipso}{erg s$^{-1}$ cm$^{-2}$ \AA$^{-1}$}

\input{psfig}

\begin{document}

\title{IC~5063: AGN driven outflow of warm and cold gas\thanks{Based on observations
with the ESO-NTT and with the Australia Telescope Compact Array}}

\titlerunning{IC 5063: AGN driven outflow of warm and cold gas}
\authorrunning{Morganti et al. }
 
\author{R. Morganti\inst{1,2}, J. Holt\inst{3}, L. Saripalli\inst{4,5},
  T. A. Oosterloo\inst{1,2}, C. N. Tadhunter\inst{3}}

\offprints{morganti@astron.nl}

\institute{Netherlands Foundation for Research in Astronomy, Postbus 2,
NL-7990 AA, Dwingeloo, The Netherlands
\and
Kapteyn Astronomical Institute, University of Groningen, Landleven 12, 9747 AD,
Groningen, NL
\and
Department of  Physics and Astronomy,
University of Sheffield, Sheffield, S7 3RH, UK
\and 
Raman Research Institute, CV Raman Avenue, Sadashivanagar, Bangalore 560080, India
\and
CSIRO, Australia Telescope National Facility, PO Box 76 Epping NSW 1710 Australia
}

\date{version $ $Date: 2007/04/13 10:15:29 $ $ }

\abstract{We present new ATCA 17- and 24-GHz radio images and ESO-NTT 
optical spectra of the radio-loud Seyfert galaxy IC~5063, the first galaxy
in which a fast ($\sim$ 600 \kms) outflow of neutral hydrogen was discovered. The
new radio data confirm the triple radio structure with a central, unresolved
flat-spectrum core and two resolved radio lobes with steep spectral
index. This implies that the previously detected fast outflow of neutral gas
is occurring off-nucleus, near a radio lobe about 0.5 kpc from the core.  The
ionised gas shows highly complex kinematics in the region co-spatial with the
radio emission.  Broad and blueshifted ($\sim 500$ \kms) emission is observed in the region
of the radio lobe, at the same location as  the blueshifted
\HI\ absorption. The velocity of the ionised outflow is similar to
the one found in \HI. The first order correspondence between the radio and
optical properties suggests that the outflow is driven by the interaction 
between the radio jet and the ISM.  However, despite the high outflow
velocities, no evidence is found for the ionisation of the gas being due to
fast shocks in the region of the outflow, indicating that  photoionisation
from the AGN is likely to be the dominant ionisation mechanism. \\
The outflow rate of the warm (ionised) gas is small compared to that of
the cold gas, similar to what is found in other radio galaxies.  The
mass outflow rate associated with the \HI\ is in the same range as 
for ``mild'' starburst-driven superwinds in ULIRGs.  However, in 
IC~5063, the AGN-driven outflow appears to be limited to the inner kpc region of the galaxy.
The kinetic power associated with the \HI\ outflow is a small fraction (a few
$\times 10^{-4}$) of the Eddington luminosity of the galaxy but is a
significant fraction ($\sim 0.1$) of the nuclear bolometric luminosity.  In
IC~5063, the observed outflows may have sufficient kinetic power to have a significant impact
on the evolution of the ISM in the host galaxy.


}
\maketitle

\begin{figure*}
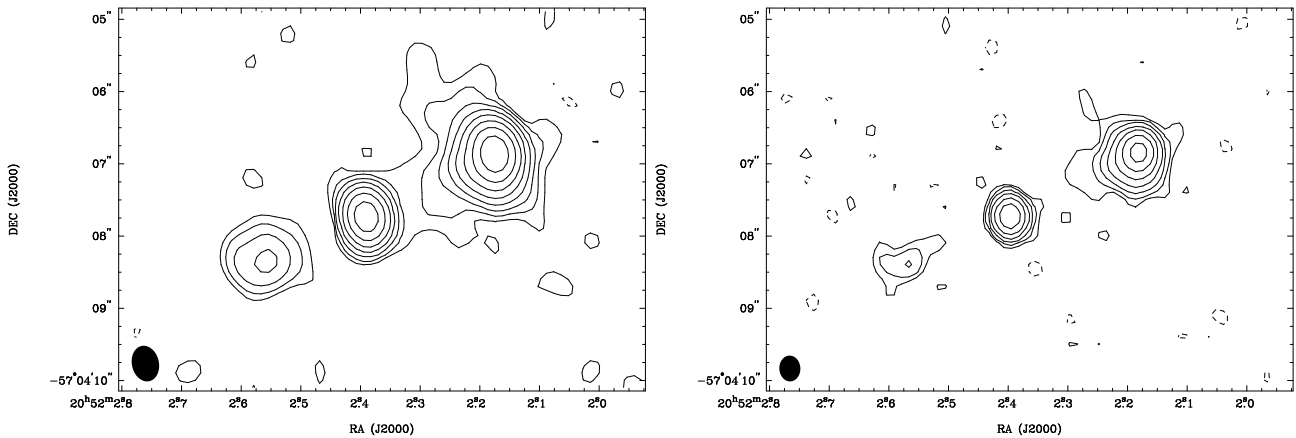

\centerline{\psfig{figure=7888Fig1a.ps,angle=-90,width=8.5cm}
            \psfig{figure=7888Fig1b.ps,angle=-90,width=8.5cm}}
\caption{The radio continuum ATCA images of IC~5063 at 17.8 GHz and 24.8 GHz. 
The contour levels are 0.28 and 0.46 $\times \hbox{-1}, 1, 2, 4, 8, 16, 32,
64, 128$ \mJybeam\ respectively.}
\end{figure*}

\section{Introduction}

Huge amounts of energy are produced through the accretion of material onto the
super-massive black hole situated in the centre of an Active Galactic Nucleus
(AGN).  This energy is released into the surrounding medium in a number of
different ways, ranging from collimated radio-plasma jets to UV and X-ray
emission.  The regions around an AGN are, therefore, highly complex and host a
wealth of physical processes.  Gas in different phases (atomic, molecular and
ionised) is observed in this very hostile environment.  This gas, in
particular through its kinematics and ionisation, can carry the signatures of
the effect of the AGN on its surrounding medium.

The energy released from the nucleus can produce gas outflows with high
velocities (thousands of \kms). This has been observed in many AGN, from
Seyfert galaxies to quasars.  Gas outflows are detected as blueshifted
absorption or emission line wings in optical, UV and X-ray spectra, (see e.g.\
Crenshaw, Kraemer \& George 2003, Krongold et al.\ 2003, Elvis et al.\ 2002,
Holt et al.\ 2006 and references therein). More recently, fast and massive
outflows of {\sl neutral hydrogen}, detected as 21-cm \HI\ absorption against
the central regions of radio-loud galaxies, have also been discovered
(Morganti, Tadhunter \& Oosterloo 2005a).

All these different types of gaseous outflows are of great interest for a
number of reasons. Understanding the driving mechanism(s) of the outflows is
crucial for understanding the physical mechanisms at work in the
central regions of galaxies with an AGN. Furthermore, massive AGN-driven
outflows are now  suggested to dramatically affect the evolution of galaxies
due to the large amounts of energy they feed back into the interstellar medium
(see e.g.\ Silk \& Rees 1998, di Matteo et al.\ 2005).

Gaseous outflows can be driven by super-winds associated with large
starbursts, (Heckman 2002, Rupke et al.\ 2005a,b; Veilleux et al.\ 2005).
However, in the case of galaxies with an AGN, radiation or wind pressure from
the regions near the active super-massive black hole (i.e.\ a quasar wind) are
the likely drivers of the gas outflows detected at X-ray and UV wavelengths.  In
radio-loud objects, the interaction of the radio plasma with the (rich)
gaseous medium in the direct vicinity of the active nucleus can provide
another mechanism that can drive the outflows (see also discussion in e.g.\
Batcheldor et al.\ 2007).

For the fast outflows of neutral hydrogen many unresolved issues remain, in
particular regarding their origin and their quantitative effect on the ISM. To
answer these questions, it is crucial to know where the outflows are occurring
with respect to the AGN, and to know the characteristics of the gas in other
phases (e.g.\ ionised). These \HI\ outflows occur, in at least some cases, at
kpc distance from the nucleus (see Morganti et al.\ 2005a). This large
distance from the AGN suggests that the interaction between the expanding
radio jets and the gaseous medium enshrouding the central regions is the
driving mechanism of such outflows.  The associated mass outflow rates can be
as large as $\sim 50$ $M_\odot$ yr$^{-1}$, comparable (albeit at the lower end
of the distribution) to the outflow rates found for starburst-driven
superwinds in Ultra Luminous IR Galaxies (ULIRG).  This would suggest that
these jet-driven outflows of neutral gas in radio-loud AGN can indeed have a
significant impact - similar to the one of superwinds - on the evolution of
the host galaxies.  It is also important to be able to compare the
characteristics of the \HI\ outflows with those of the ionised gas.  For
example, in the radio galaxies 3C~305 and 3C~293 (Morganti et al.\ 2005b,
Emonts et al.\ 2005, Morganti et al.\ 2003) the mass of the gas in the ionised
outflow is much less than that in the neutral outflow, making the neutral
hydrogen the dominant and therefore the most influencing component in the
feedback process.

\begin{figure}
\centerline{\psfig{figure=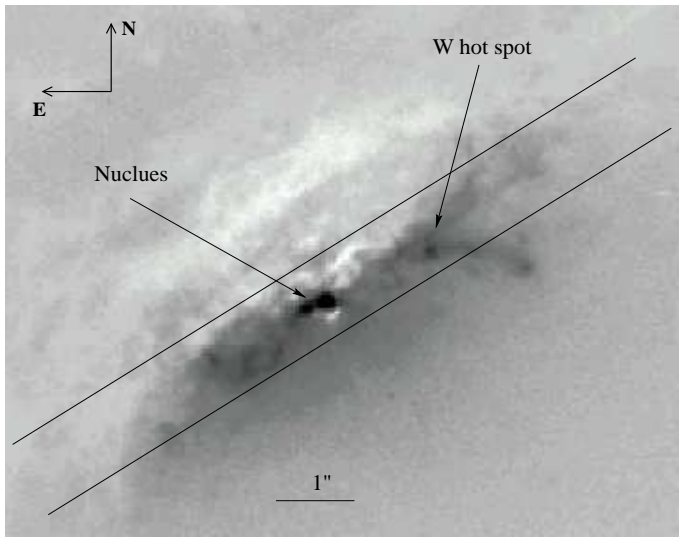,width=9cm,angle=0}}
\caption{The slit position of the NTT observations superimposed on the HST
image (from the public archive) obtained with WFPC2 through filter F606W (that
includes bright emission lines like [\OIII]).}
\label{fig:slits}
\end{figure}

All these considerations motivated a new study of the radio-loud Seyfert 
galaxy IC~5063.  IC~5063 ($z = 0.0110$) is classified as a Seyfert 2 and is
classified as an early-type galaxy that shows a complicated system of
dust-lanes (see Figure 1). 
In polarized light IC~5063 shows high polarization in the near IR (Hough et
al.\ 1987) and a strong, broad \ha\ emission (Inglis et al.\ 1993).  Like in
some other Seyfert 2 galaxies this suggest that there is a broad-line region
which is obscured from our direct view and the broad-line radiation is
scattered into our line of sight by scatterers outside the obscuring regions.
IC~5063 is among the most radio-loud Seyfert galaxies known, while its radio power
is at the lower end of the distribution for radio galaxies. IC 5063 is the
first galaxy where a fast neutral outflow was discovered (Morganti et al.\
1998).  In the radio continuum, IC 5063 shows a triple structure of about 4
arcsec in size (about 1.3 kpc\footnote{We assume $H_\circ= 71$ \kmsMp\ which
implies a distance of 46.5 Mpc; 1 arcsec is equivalent to 0.22 kpc }; see
Morganti et al.\ 1998)  aligned with the dust-lane. The neutral outflow occurs against the NW and
strongest radio component (Oosterloo et al.\ 2000; see also below).  With new
high-resolution and high-frequency radio observations obtained with the
Australia Telescope Compact Array (ATCA) and deep long-slit optical spectra
obtained with the ESO New-Technology Telescope (NTT), we aim to study the
optical kinematics and ionisation properties in relation to the radio
structure.  The kinematics (and the ionisation level) of the gas shed light on
whether a strong jet/cloud interaction is responsible for the outflow or
whether other processes must be invoked. This object also allows a detailed
comparison between the properties of the ionised- and neutral-gas outflows.

\begin{figure}
\centerline{\psfig{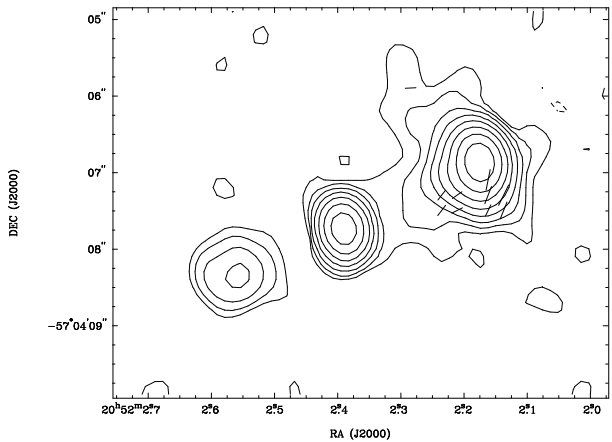}}
\caption{Vectors proportional to the 17~GHz fractional polarisation (with a 3$\sigma$
cut-off) superimposed on the continuum 17~GHz image. Vector position angle has been corrected for 
Faraday rotation.
}
\end{figure}

\section{The radio structure: ATCA observations}

IC~5063 was observed with the ATCA using the 6-km configuration - the longest
available for this radio telescope.  The observations were carried out on 11
May and 8 July 2004 at 17.8 and 24.8 GHz respectively. These high frequencies
allow us to achieve a relatively high resolution.  We took data simultaneously
at 17.73 and 17.86 GHz in the first observation and at 24.77 and 24.90 GHz in
the second, with a bandwidth of 128 MHz for each of these frequencies.  These
separate frequencies allowed us to improve (radially) the $uv$ coverage.  For
both observations the total integration time on IC~5063 is about 9~h
consisting of many cuts of 10 min spread over a 12-h period.  PKS~1934--638
was used as the primary (flux) calibrator (with an assumed flux of 1.079 Jy at
17.8~GHz and 0.745 Jy at 24.8~GHz).  A nearby secondary calibrator (PKS
2052--474) was observed for 1 minute every 10 minutes as well as for reference
pointing. This reference pointing was done every hour.

The data reduction was done using the MIRIAD package (Sault, Teuben \& Write
1995). The final images were obtained using a few cycles of self-calibration
and using uniform weighting to achieve the highest possible resolution.  The
17-GHz image has a beam size of $0.5 \times 0.37^{\prime\prime}$ (12$^\circ$)
and an rms noise of 0.11 \mJybeam. At 24 GHz the beam is $0.36
\times 0.29^{\prime\prime}$ (2.5$^\circ$) with an rms noise of 0.18 \mJybeam.

Fig.\ 2 shows the 17- and 24-GHz continuum images. As expected from the
previous 8-GHz observations (Morganti et al.\ 1998), the source shows a
triple structure oriented along P.A.$\sim 115^\circ$. In contrast to previous
lower resolution images, the bright NW source appears to be connected to the
core by a low-brightness bridge.  Although most of the flux in the NW source
appears to come from a  quite compact and bright region,
low-brightness, diffuse emission surrounding the NW source is also detected.

The new, higher-resolution observations allow us to investigate in more detail
the structure of the radio emission while, due to the broad frequency range of
the new observations, we are able to accurately derive the spectral index.
The spectral index is crucial information for identifying which of the
features correspond to the core and which to radio lobes.

The total fluxes of the three components at the two different frequencies are
listed in Table{\ref{tab:radio}}.  The corresponding spectral indices
($\alpha$ defined as $S = \nu^{-\alpha}$) are also listed in this table.  In
contrast to the E and W component, the central component has a flat spectral
index ($\alpha = 0.2$), confirming the earlier prediction that this component
is the radio core in IC~5063 (Oosterloo et al.\ 2000).  The two east and west
extended components show, instead, much steeper spectra, confirming their
nature as radio lobes.  Thus, the new observations support the interpretation
of the radio structure given by Morganti et al.\ (1998). Given that the
neutral hydrogen outflow is detected (via VLBI observations, see Oosterloo et
al.\ 2000) against the stronger (i.e.\ the western) radio component, the new
observations now confirm that this neutral outflow is occurring against {\sl a
lobe located about 0.5 kpc from the core}.

We have derived the polarisation characteristics of the source at the two
frequencies. We have obtained images of the Stokes parameters ($Q, U$), the
polarised intensity ($P=(Q^2+U^2)^{1/2}$) and the position-angle of the
polarisation ($\chi=0.5 {\rm arctan} (U/Q)$). The rms noise of the $Q$ and $U$
images is about 0.08 and 0.1 \mJybeam\ at 17 and 24 GHz respectively.

The polarised intensity and the fractional polarisation ($m=P/I$) were
estimated only for those pixels for which $P>3\sigma_{QU}$.
Polarised emission is found mainly over an elongated feature in the western radio
lobe, stretching from the total intensity peak to the SW region of the
lobe. Fractional polarisation values between 10 -- 20\% are seen towards the
SW region at the two frequencies. The fractional polarisation decreases
steadily towards the NE with the lowest values of under 1\% measured
in the centre of the
lobe. Polarised flux is detected only in the western lobe.

\begin{table} 
\centering 
\begin{tabular}{lccccc} 
\hline\hline\\ 
Region &  17~GHz     &  24~GHz    & $\alpha$ \\
       &  mJy        &  mJy       &     \\
\hline
       &             &            &          \\
   E   &    5.6      &  4.7       &  0.53$\pm0.02$   \\
  core &  30.8       &  29.0      &  0.18$\pm 0.02$   \\
   W   &  91.2       &  67.0      &  0.93$\pm 0.02$  \\
\hline\hline
\end{tabular}
\caption{Flux densities and spectral indices for the radio three components in IC~5063 }
\label{tab:radio}
\end{table}

In Fig.\ 3 we have overlayed the position angle distribution of the electric
vectors corrected for Faraday rotation on the 17~GHz total intensity
map. Using the values of the polarisation position angle at the two
frequencies, we derive the rotation measure ($RM$) over the western lobe. We
obtain high $RM$ values over the region of up to a few thousand rad m$^{-2}$:
at $5\sigma$ level the $RM$ ranges between 2000 and 3000 rad m$^{-2}$.
We do not think the $RM$ derivation
suffers from $n\pi$ ambiguities. At these high frequencies, the $RM$ required
for a rotation in position angle of $\pi$ radians is about 140000 rad m$^{-2}$
- a very large and unlikely value.


\begin{table*}
\centering
\begin{tabular}{lccccccccc}\hline
Date  & Arm & Exposure & Set-up & Slit PA & Slit width & $\lambda_c$ &
$\Delta\lambda$&Seeing$\dagger$  & Conditions \\
yyyy/mm/dd & &   s & CCD+grating & degrees & arcsec & \AA\ & \AA\ &
arcsec & \\\hline
2000/07/25  & B & 4$\times$1800 & TK1034+\#12 & 115 & 0.8 & 4052 &
3560-4460 & 0.8-1.0& Photometric\\
2000/07/25  & R & 2$\times$1800 & TK2048+\#7 &  115 & 0.8 & 5150 &
4450-5760 & 0.8-1.0& Photometric\\
2000/07/25  & R & 2$\times$1800 & TK2048+\#7 &  115 & 0.8 & 6351 &
5700-7000 & 0.8-1.0& Photometric\\
\hline
\hline
\end{tabular}
\caption[]{Log of observations.  Seeing is estimated using the DIMM
seeing measurements and service observer's log.} 
\label{tab:obs}
\end{table*}

\begin{figure*}
\centerline{\psfig{figure=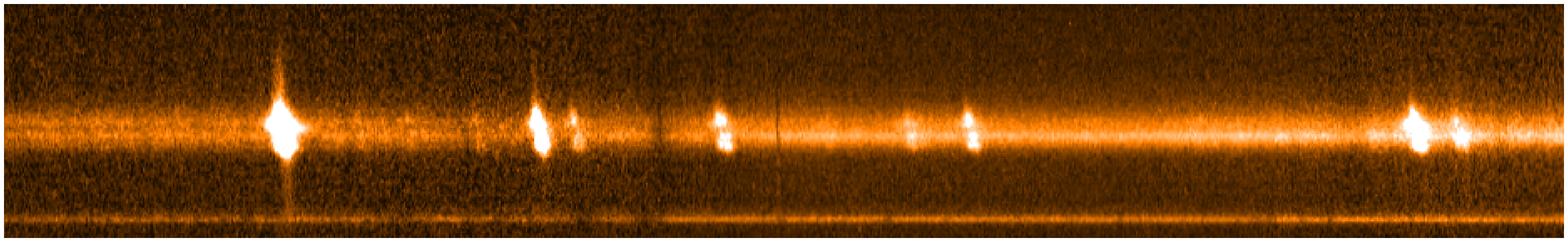,width=18cm,angle=0}}
\centerline{\psfig{figure=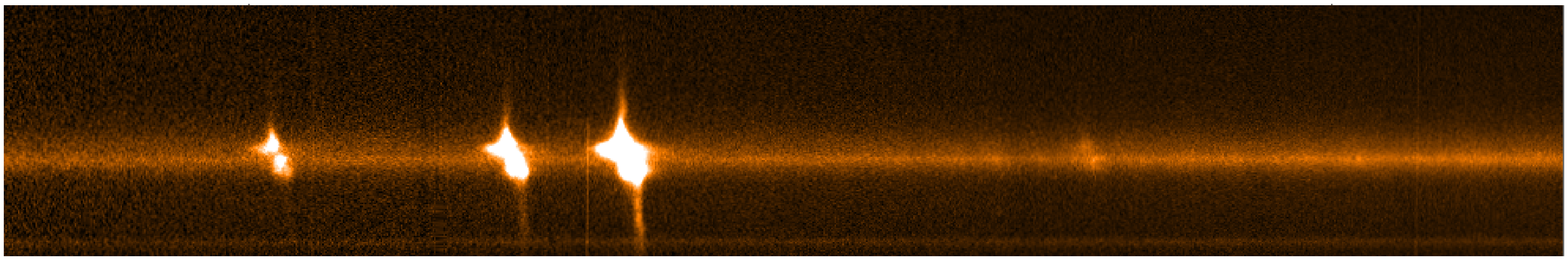,width=18cm,angle=0}}
\centerline{\psfig{figure=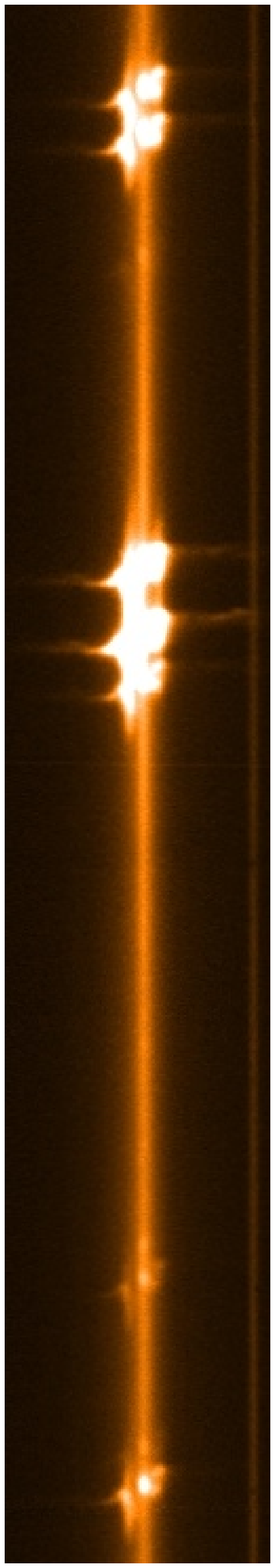,width=18cm,angle=-90}}
\caption{The spectral regions from {[\OII]}\lala3727 to {[\OIII]}$\lambda$4363
(top), H$\beta\lambda$4860 to {[N I]}$\lambda$5199 (middle) and
{[OI]}\lala6300,6363 to [\SII]\lala6716,6731 (bottom) along PA 115$^{\circ}$. North-west
is to the top, south-east to the bottom.
Note, the spectra shown do not have identical scaling in either flux or
spatial scale.  Rather, the scales are optimised to highlight  the differences in the
intensity of the lines and allows the different kinematic components to be
identified: from regular galaxy rotation at larger radii to the complex
kinematics in the nuclear regions, clearly observed in all lines and is
highlighted particularly well in the {[\OI]}\lala6300,6363 lines in this
Figure.}
\label{fig:2dspec}
\end{figure*}

\section{The ionised gas: NTT observations}

In order to obtain more information about the conditions of the gas in the
inner regions of IC 5063, we have
investigated the characteristics of the ionised gas using long-slit spectra
obtained - in service mode - with the EMMI spectrograph on the 3.5-m
New Technology Telescope (NTT) in La Silla, Chile. Using EMMI in medium
resolution mode, we obtained simultaneous spectra in the red and in the blue
arm.
The observations are summarised in Tab.\ 2.
In the blue, one central wavelength setting was used ($\lambda_c$ = 4052
\AA)  to obtain spectra covering 3560-4460 \AA.   In
the red arm, two settings for the central wavelength, $\lambda_c$ = 5150 \AA\
and $\lambda_c$ = 6315 \AA, were used yielding spectra with useful wavelength
ranges of 4450-5760 \AA\ and 5700-7000 \AA\ respectively.  Spectra were taken
along PA 115$^{\circ}$ (see Figure \ref{fig:slits}) with a 0.8-arcsec
slit. This PA lies along the major axis of the radio emission as well as the
dust-lane and includes both the nucleus and the radio hotspot, $\sim$2 arcsec
to the west. The spatial pixel scale is 0.27 arcsec/pix. To reduce the effects
of differential refraction, all exposures were taken when IC~5063 was at low
airmass (sec $z$ $<$ 0.1).

The data were reduced in the usual way (bias subtraction, flat fielding,
cosmic ray removal, wavelength calibration, flux calibration) using the
standard packages in { IRAF}.  The two-dimensional spectra were also corrected
for spatial distortions of the CCD.  The final wavelength calibration
accuracy, calculated from the locations of the night sky emission lines
(Osterbrock et al.\ 1996) is between 0.05 and 0.1 \AA\ for the red
spectra. The spectral resolution estimated from the widths of the night sky
emission lines is 2.0 - 2.4 \AA\ for the red spectra. Due to the absence of
sky lines in the blue spectra, we have not been able to quantify the
wavelength calibration accuracy nor the spectral resolution in the blue, but
there is no reason to believe that it is very different from that in the red
spectra.

Comparison of the data of several spectrophotometric standard stars taken with a
wide slit (5 arcsec) throughout the run gives  a relative flux
calibration accuracy of about 5 per cent. This accuracy is confirmed
by the good matching of the continuum flux for the different regions.

\begin{figure*}
\centerline{\psfig{figure=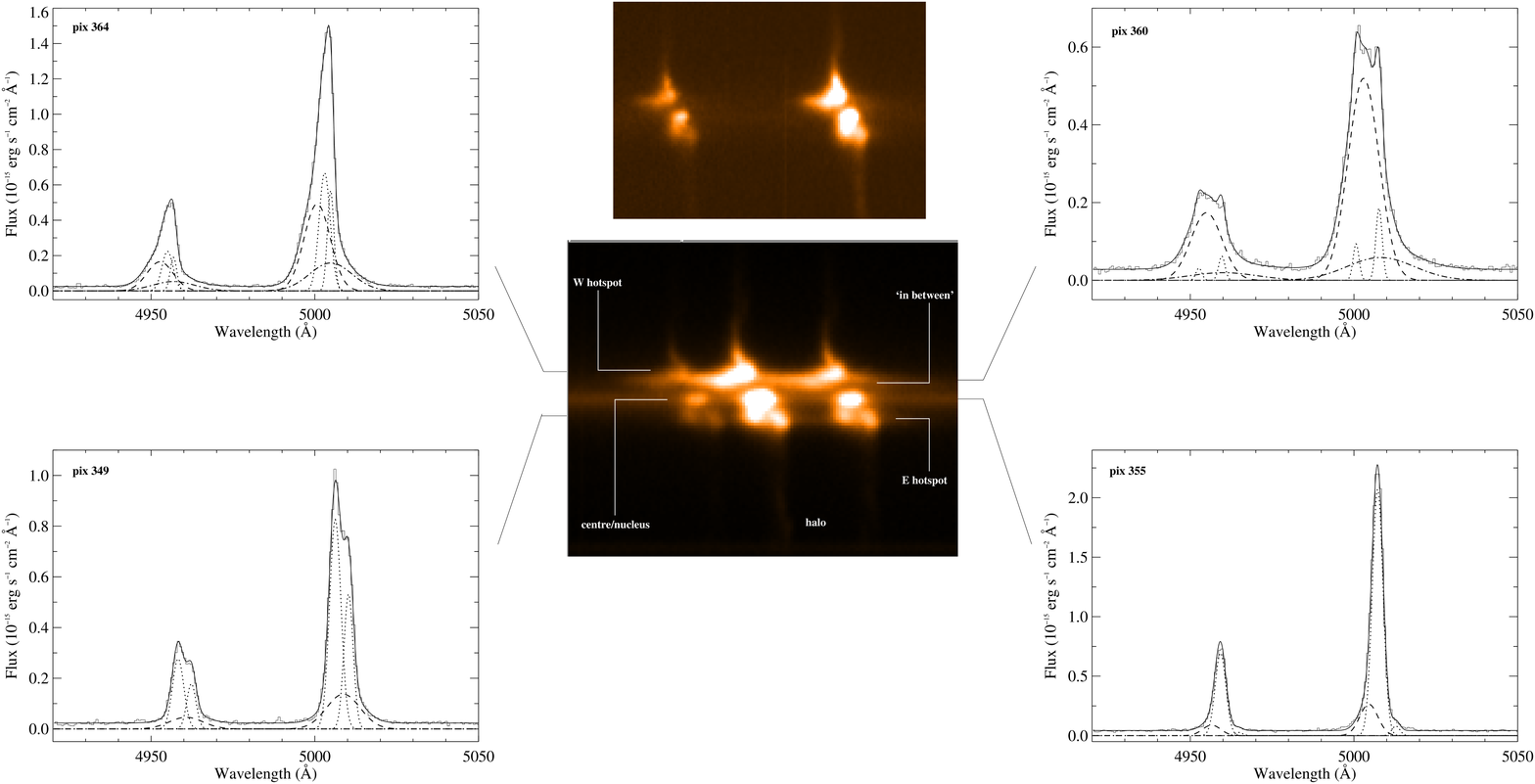,width=18cm,angle=00}}
\caption{Spatial variations of the emission line profiles. In the centre 
  is the 2D spectrum, zoomed in on the {[\OIII]}\lala4959,5007 (top)
H$\alpha$/\NII\ blend (bottom). The four regions
  discussed in Section 4 are identified. On the left and right, a selection of
  the emission line profiles of {[\OIII]}\lala4959,5007 at various spatial
  positions are shown along with the best fitting model and the various
  sub-components required to model the lines. Key: black line (data), thick
  line (overall model) dashed lines are the different Gaussian used in the
  fit, see text for details. }
\label{fig:2Dhalpha}
\end{figure*}
\begin{figure*}
\centerline{\psfig{figure=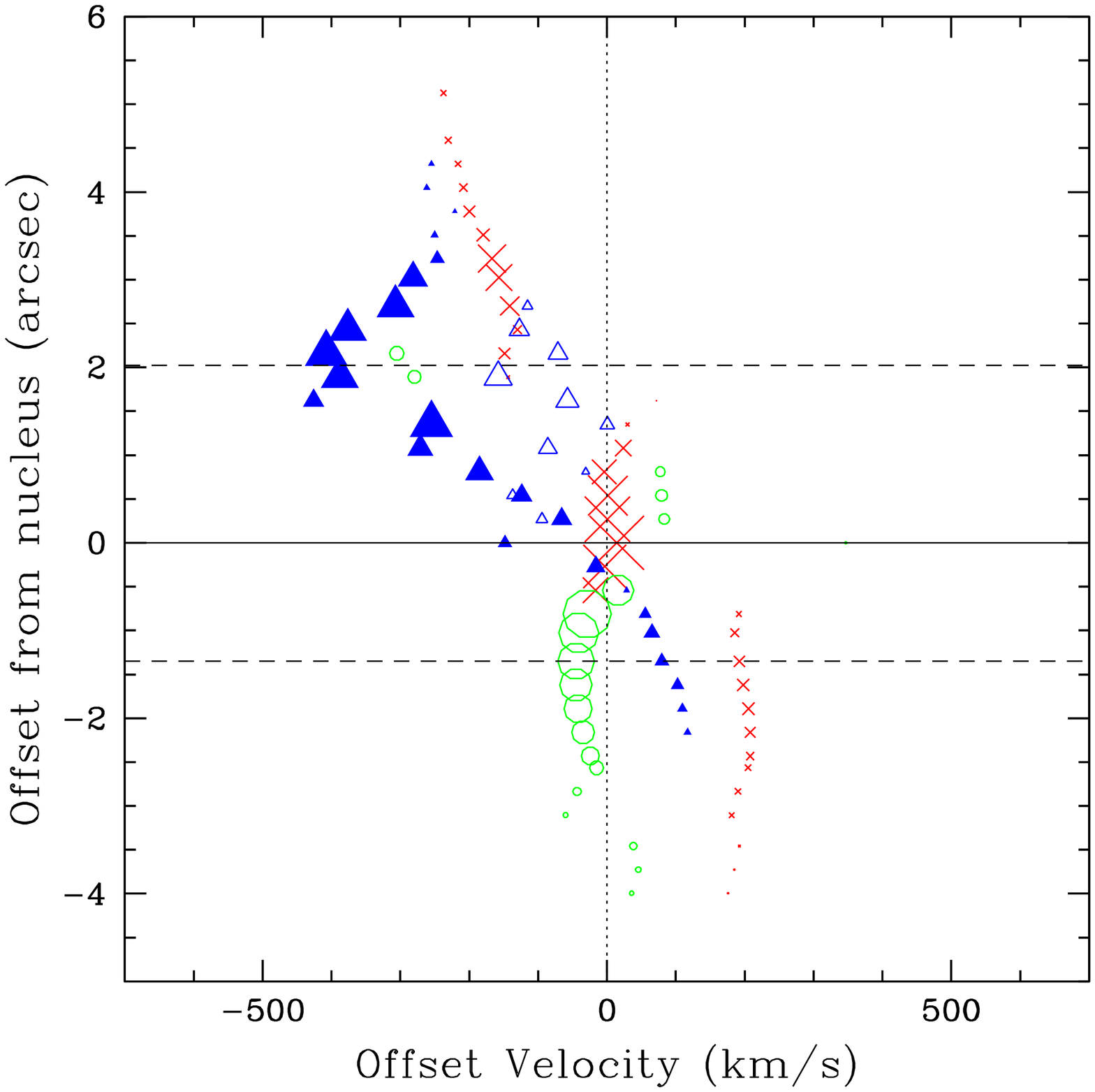,width=7.5cm}
\psfig{figure=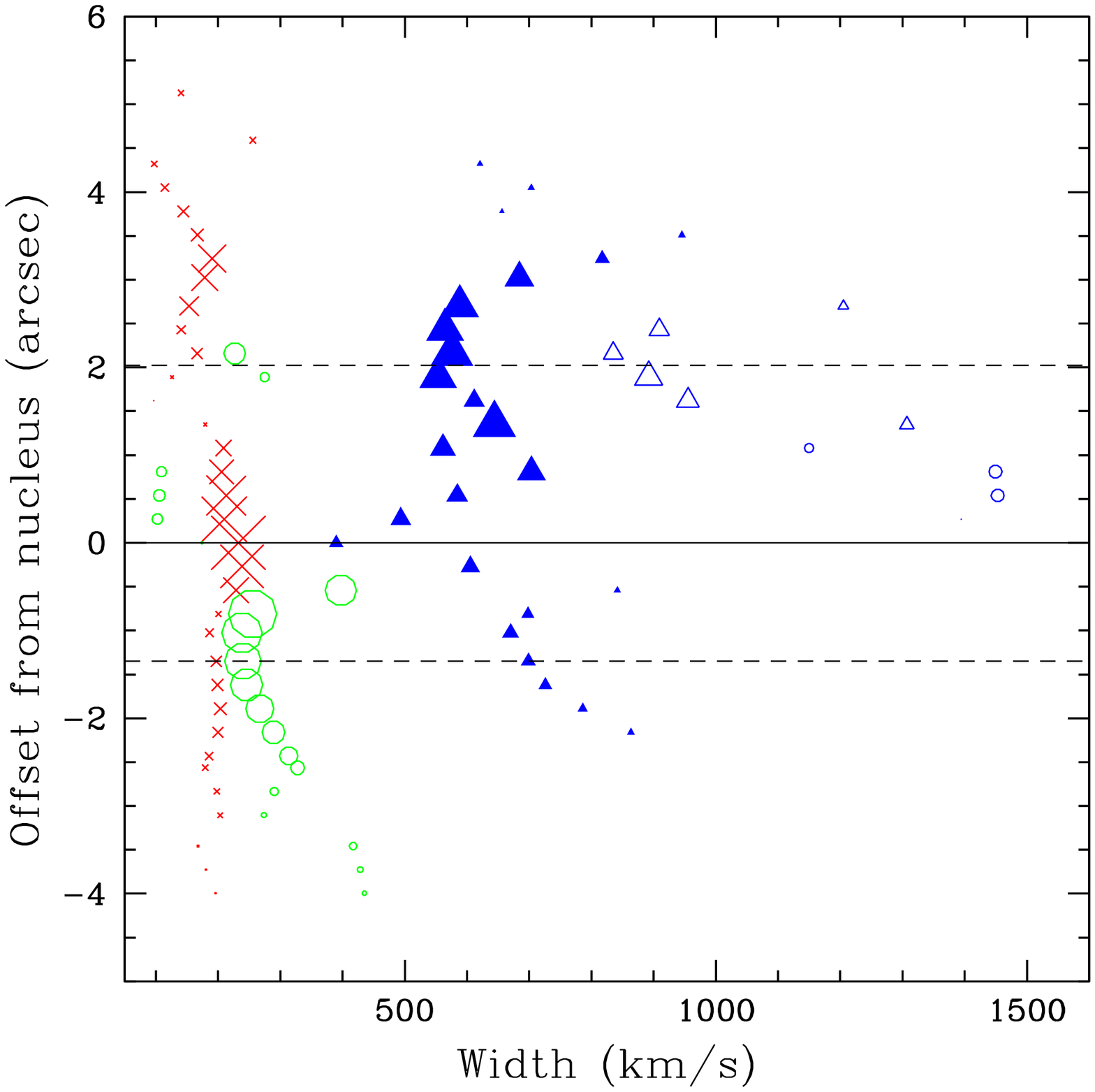,width=7.5cm}}
\caption{Spatial variations of the velocity width and shift of the
  various kinematic sub-components as measured from the
  {[\OIII]}\lala4959,5007 doublet. Crosses (red, in the electronic version) and open circles
(green)  represent the narrow components, filled triangles represent the extended
  broad component and  open triangles (blue) represent the extremely broad
  component observed in the western hotspot. In addition, the points are
  scaled according to the measured line flux in the components.  The solid horizontal
  line marks the position of the nucleus, and the dotted lines mark the
  position of the western (top) and eastern (bottom) hotspots.  The systemic
  velocity used to estimate the ``Offset velocity'' is 3400 \kms\ as derived
  from
\HI\ observations (Morganti et al. 1998).
  }
\label{fig:vwo3}
\end{figure*}

\subsection{Line fitting and ionised gas kinematics}

Figure {\ref{fig:2dspec}} shows the 2-dimensional spectra of IC~5063.  Due to
the very good spatial resolution and the high S/N, the 2-dimensional spectra
were analysed on a pixel-by-pixel, except in the outer regions where the
lower S/N necessitated binning. 

Before proceeding, we would like to stress that  
in this paper we are mainly going  to use the
red spectrum and the analysis will focus on the kinematics of the
emission lines. The main reason for this is that a proper analysis of the line
ratios and diagnostic diagram would require a proper fitting of the underlying
stellar continuum (as done in the case of both PKS~1345-12 and PKS~1549-79,
Holt et al. 2003, 2006). This is important in order to take into account
possible stellar absorption lines (if a young or intermediate stellar
population is present, something that cannot be excluded for 
IC~5063 given the gas rich medium). A first-order discussion of the emission
lines ratios is given in Sec. 3.2 while a full analysis is postponed to a
future paper.

The spectra were extracted and analysed using the {STARLINK} packages {FIGARO}
and { DIPSO}.  Following the methods of Holt et al.\ (2003), we have modelled
the emission lines, in each aperture, using multiple Gaussian components.  We
initially modelled the strong [\OIII]\lala4959,5007 doublet and defined a
`good fit' as that comprising the minimum number of Gaussian components to
provide a physically viable good fit {\sl at each position}.  We will refer to
these models as the `{[\OIII]} model' hereafter.  The offset velocity is
defined relative to the systemic velocity ($\sim 3400$ \kms) derived from the
\HI\ observations (Morganti et al.\ 1998).  Up to four Gaussian components
were required in the outflowing region while two or more components were
required in other regions.  Interestingly, the central region -- corresponding
to the AGN nucleus -- is the least kinematically disturbed region.  Examples
of the [\OIII] emission line profiles in different regions are shown in Figure
{\ref{fig:2Dhalpha}}.  The location of the radio lobe is also indicated for
direct comparison between the characteristics of the ionised gas and the
location of the radio emission.

The spatial variations of the velocity widths and shifts and intensity of
these components are shown in Figure {\ref{fig:vwo3}}.  We have attempted to
group the various components using their kinematical properties to aid in the
description below.  We realise that the separation in Gaussian components, as
well as grouping them into ``kinematical components'', can be somewhat
artificial and subjective.  We have based our grouping on {\sl kinematical
continuity} and/or in similarities (e.g.\ FWHM) of the components.  The
kinematics of the ionised gas are complex, in particular near the NW radio
lobe.  We identify the following main kinematic components:

\begin{enumerate}
\item 
As can be seen in Figures {\ref{fig:2dspec}} and {\ref{fig:2Dhalpha}}, the
emission lines are highly extended along PA 115$^\circ$.  The brightest lines
(H$\alpha$/{[\NII]}, {[\OIII]}\lala4959,5007, {[\OII]}\lala3727 and 
{[\SII]}\lala6716/6731) are observed up to 15.1 arcsec (3.3 kpc) to the E and 16.2
arcsec (3.6 kpc) to the W of the core. At these distances from the centre, a {\sl narrow
component} (FWHM $\sim$ 100-200 \kms), which follows the rotation of the
galaxy with velocity amplitude $\sim$ 200 \kms\ is detected (the crosses in
Fig.\  6).

\item In the region from the nucleus  extending to at least 4 arcsec SE and 2 arcsec NW,
we detect a second narrow component (FWHM $\sim$ 200-300 \kms, the open
circles in Fig.\ 6) close to the systemic velocity.  This component is seen
outside the region of radio emission on the E side.

\item A {\sl broad component}, 500 $<$ FWHM $<$ 700 \kms,
is observed along the entire spatial extent of the radio emission (the filled
triangles in Fig.\ 6). In the
region around the NW lobe (where we see the most extreme kinematics), this
broad component appears highly {\sl blueshifted} ($\sim 500$
\kms), clearly deviating from the regular kinematics of the galaxy.

\item A {\sl very  broad component}, (up to $\sim$ 900-1300 \kms) 
only present around the region coincident with the NW radio lobe, and 
centred on the systemic velocity (i.e.\ not blueshifted, open triangles in
Fig.\ 6)

\end{enumerate} 

Hence, it is clear that the kinematics of the ionised gas is very
complex. In particular, we note that there is a close correspondence between
the ionised gas kinematics and the radio structure while the most extreme
ionised gas kinematics occur, to first order, at the location where the fast outflow
of neutral gas is detected. All these features suggest that {\sl a strong
interaction between the radio plasma and the ISM} is taking place.

\subsection{Physical conditions of the ionised gas}

A few interesting features can be extracted from the various line ratios
observed. We will concentrate  on a few relevant regions (see
Table {\ref{tab:lines}}).

In the nuclear region,  many strong emission
lines are observed, as previously reported by Colina, Sparks \& Macchetto
(1991).  We confirm the detection of high-ionisation species such as
{[\FeX]}$\lambda$6375, {[\FeVII]}$\lambda$5721 and {[\FeVII]}$\lambda$6087.
In addition, we observe a feature at 5309\AA, similar to that observed in PKS
2152--699 by Tadhunter et al.\ (1989) which we identify as
{[\CaV]}$\lambda$5309.
The region of the SE radio lobe is also characterised by strong,
high-ionisation emission lines.  In the region of the NW lobe, many bright
emission lines are observed, although the highest-ionisation species found in
the nucleus and eastern radio hotspot are not observed in this region.

Optical images of IC~5063 show a complex dust lane system (see  Figure
{\ref{fig:slits}}) and, as our slit was aligned along this dust lane, it is
likely that some of the emission line components will be reddened to some
degree.  We have estimated the degree of reddening in three regions (nucleus,
northern outflowing region and the southern region with double narrow lines)
using the commonly used Balmer decrement \ha/\hb. In all regions and in all
kinematic components, we measure significant reddening, typically \ha/\hb~$>$
6, corresponding to $\ebv> 0.5$. In all regions, the largest reddening is
found in the component with intermediate blueshift and in one of the two narrow
components (see Table\ {\ref{tab:lines}}).  Whilst our results are consistent
with the presence of a large, patchy dust lane, the specific numbers must be
used with caution.  No detailed modelling of the optical continuum was done and
therefore no corrections for stellar absorption lines were applied to e.g.\
the \hb\ flux and  it is therefore possible that we have underestimated the
\hb~flux due to underlying stellar absorption, and so overestimated the degree
of reddening.   We have performed a quick fit to the stellar continuum to
estimate the worst case scenario for possible underlying \hb~absorption. In
the southern and northern apertures, the effect of stellar absorption lines is
negligible to a few percent and our estimated uncertainties on the emission
line fluxes will cover this. In the northern region, the effect is again up to
a few percent in the narrow components. However, the effect is perhaps as
large as 20-25\% in the intermediate component in the central region. Keeping
this uncertainty in mind, we can make a few remarks about the line ratios.

After modelling {[\OIII]}\lala4959,5007 in the various regions (see Sect.\
3.1), we have attempted to model some of the other emission lines  (limited to
\hb\ and [\SII]6716/6731) {\sl using the same
kinematic model as derived for the [\OIII] for the same aperture}. We fixed
the velocity widths and shifts and only allowed the relative fluxes of the
kinematic subcomponents to vary. 

For one of the narrow components (the circles in Fig.\ 6), the
[\OIII]/H$\beta$ ratio is high ($\sim 12$) in the centre.  As this component
is narrow, it is unlikely to be strongly influenced by the interaction of the
radio plasma with the ISM and the ionisation of this gas is likely due to the
{\sl UV} radiation from the nucleus.

High {[\OIII]}/\hb\ ratios (between 10 and 12) are also observed in the narrow
components at the location of the radio lobes.  Furthermore, high values of
the [\OIII]/H$\beta$ ratio are also observed in the intermediate-width (600
\kms), blueshifted component in the lobes.  Even if no detailed analysis of
the line ratios is done in this paper for IC~5063, the {[\OIII]}/\hb\ ratios
allow a useful comparison with the detailed studies of jet-cloud interactions
in radio galaxies.  In the radio galaxies the {[\OIII]}/\hb\ ratios provide
the best evidence for shock ionization: the shocked components tend to have
low {[\OIII]}/\hb\ ratios (see e.g.\  3C~171: Clark et al.\ 1997; PKS 2250--41:
Villar-Martin et al.\ 1999), in contrast to the situation of the kinematically
disturbed components to to the NW in IC~5063.  Thus, this suggests that even
if these components would originate from gas shocked by the interaction
between the radio plasma and the ISM, these shocks are not the dominant
ionisation mechanism.  Although this result will need to be confirmed using
more emission lines (and after a proper continuum subtraction), it
nevertheless supports the conclusions based on the energy budget (Morganti et
al.\ 1998) and is also consistent with what has been found in similar studies
of other Seyfert galaxies, e.g.\ Mrk 78 (Whittle et al.\ 2005).

\begin{table*}
\centering

\begin{tabular}{llcccccc}
\hline\hline\\
Region       & component  & H$\beta$ flux & [\OIII]/H$\beta$ & H$\alpha$/H$\beta$ &
[\SII]6716/6731  & density   & \ebv  \\
              &               &  10$^{-16}$   &          &             & & 
& \\
              &               & \dipso   &          &             & & 
cm$^{-3}$  & \\
\hline
              &               &                &      &  & & \\
   { SE} & narrow ($\times$)  &  $7.8\pm1.2$  &
   $9.3\pm0.2$ & $10.1\pm0.2$
& $1.04\pm0.05$    &   513$\pm$65   & 0.94 \\
              & narrow ($\circ$)   &  $9.3\pm1.2$   & $12.8\pm0.1$ & 
$11.1\pm0.1$
& $1.06\pm0.05$ &   471$\pm$60  & 1.02 \\
              & broad ($\blacktriangle$)   &  $11.6\pm3.2$  & 
$9.4\pm0.1$  & $8.1\pm0.3$ &
$1.1\pm0.06$    &   395$\pm$100   & 0.77\\
   { centre}& narrow ($\circ$) & $31.0\pm2$    & $12.1\pm0.1$ & 
$12.9\pm0.1$ &
$0.99\pm0.05$     & 632$\pm$80   & 1.14  \\
             & narrow ($\times$)     &  $2.8\pm1.5$   & $4.6\pm0.5$ & 
$8.7\pm0.5$ &
$0.99\pm0.19$    & 632$^{+750}_{-365}$    &0.82  \\
             & broad ($\blacktriangle$) &  ($4.1\pm2.5$)   & 
($31\pm1$) & ($26\pm1$) & $0.94\pm0.14$
& 774$^{+610}_{-340}$    &1.70   \\
  { NW 1 } & narrow ($\times$)   & $0.9\pm0.6$ & $5.8\pm 0.6$ 
& $1.0\pm 0.6$ &
1.42 & $<100$ &0.0 \\
                 &  narrow ($\circ$)  & $1.7\pm0.4$ &  $8.9\pm0.2$ & 
$4.5\pm0.3$ &
1.42 & $<100$ & 0.30\\
                 & broad ($\blacktriangle$) & $5.0\pm0.9$ & $9.7\pm 
0.2$ & $7.5\pm0.2$ &
$0.5\pm0.17$ & $>3000$  & 0.70 \\
                 & very broad ($\triangle$)    &  $ 5.2\pm0.9$ &  $6.9\pm0.2$ & 
$8.1\pm0.2$ &
$1.1\pm0.6$& 395$\pm$52 & 0.77      \\
{ NW 2} & narrow ($\times$) &  $2.8\pm0.2$   &  $6.3\pm0.1$ 
& $3.5\pm 0.1$ &
1.42 & $<100$   & 0.34  \\
                 & narrow ($\circ$)  & $1.2\pm0.2$  &   $12.8\pm0.2$ 
& $7.6\pm0.2$ &
1.42 & $<100$ & 0.72 \\
                 & broad ($\blacktriangle$) & $4.2\pm0.4$  & 
$13.3\pm0.1$ & $9.2\pm0.1$
& $0.7\pm 0.1$ &  $2390\pm { 1000}$  & 0.87 \\
                 & very broad ($\triangle$)    &   $6.4\pm 0.4$ & $4.0\pm0.1$ & 
$6.1\pm0.1$
& $1.22\pm0.06$ & 213$\pm$25    & 0.54           \\
\hline\hline
\end{tabular}

\caption{Emission lines measured in three main regions.  Symbols are as in Fig.\ 6.  The errors listed here are the
combination of those associated with the fitting and due to flux calibration
error.  In bracket are very uncertain values. The fluxes for the SE region are
derived from a 0.8 arcsec  wide aperture. The NW represents two locations
bracketing the hot spot.}
\label{tab:lines}
\end{table*}

As with all other strong emission lines in IC~5063, the density diagnostic,
the [\SII]$\lambda\lambda$6716,6731 emission line doublet, is spatially
resolved enabling us to investigate the density distribution of the
emission-line gas. By modelling the doublet with the {[\OIII]} model, we
derived the density sensitive ratio 6716/6731 \AA\ in the three regions
mentioned earlier (Table\ 3).  The narrow component seems to show a density of
at most few hundred particles per cubic centimeter. For the NW region of the
fast outflow, the best fit was obtained by setting the weak narrow components
to the low density limit.  The intermediate-width ($\sim 600$ \kms), blueshifted
component observed in the NW region (as well as the intermediate component in
the SE lobe) shows a much larger density ($> 2000$ cm$^{-3}$). This is
consistent with what is expected if these broad line are produced in the
region of interaction between the radio plasma and the surrounding medium.

\begin{figure}
\centerline{\psfig{file=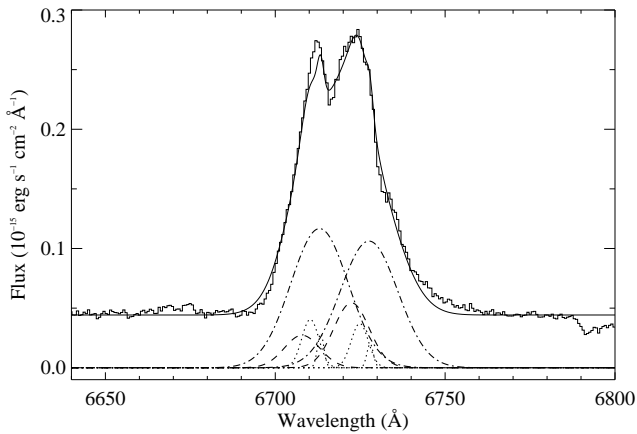,width=8.5cm,angle=0}}
\caption[]{Example of fitting of the [\SII] lines for
the broad, blueshifted component 2 arcsec NW of the nucleus superimposed onto
the data. The black line
represents the overall model, the dotted line the narrow component, the
dashed line the intermediate and the dot-dashed the broad component.}
\label{fig:s2}
\end{figure}

\section{Discussion}

The observations presented in this paper give strong evidence for the idea
that the radio plasma ejected by the Seyfert nucleus in IC~5063 is strongly
interacting with the ISM of this galaxy.  The new, high-frequency radio
observations demonstrate that the brightest radio component is indeed a radio
lobe. This implies that the fast, outflow of neutral gas is occurring
off-nucleus, about 0.5 kpc from the core. The deep long-slit optical spectra
show that the kinematics of the ionised gas is very complex over the entire
region in which the radio emission is observed. The most extreme kinematics, in
the form of a fast outflow of ionised gas, is detected at exactly the same
location as  the neutral outflow detected in \HI\ 21cm. On the other hand, despite
the complex kinematics, the gas is most likely photoionised by the Seyfert
nucleus.  We now will discuss various aspects of these observations is some
more detail.

\subsection {Radio continuum and \HI\ outflow}

Our new radio observations confirm the triple structure of the Seyfert galaxy
IC~5063. From the estimated spectral index, it is clear that the central
component corresponds to the radio core, while the other two components are
radio lobes with a steep spectral index. The bright NW radio lobe is connected
to the core by a faint bridge of radio emission that  perhaps is a 
radio jet.  The broad blueshifted \HI\ absorption found in previous \HI\
observations (Morganti et al.\ 1998) is located -- as determined by  VLBI observations --
against this NW radio lobe (Oosterloo et al.\ 2000). Our new radio
data therefore confirm that the outflow is occurring off-nucleus, i.e.\ against
a radio lobe about 0.5 kpc from the nucleus. In this respect, IC~5063 is
similar to other cases studied like 3C~305 and 3C~293 (Morganti et al.\ 2005b,
Morganti et al.\ 2003 and Emonts et al.\ 2005).

The location of the fast outflow suggests that it is more likely caused by the
interaction between the radio jet and the ISM than by other mechanisms such as
e.g.\ radiation pressure. 
The interaction scenario assumes that a particularly rich medium is present
around the NW radio lobe. That this is indeed the case is suggested by the
detection, using NICMOS, of an asymmetric H$_2$ emission, much stronger near the western lobe
(Kulkarni et al.\ 1998)  suggesting that an excess  of
molecular gas is present on the western side.  Such emission is usually interpreted as evidence for
fast shocks   (for example, if the radio jet
has struck a molecular cloud). 
The large rotation measures found in the W lobe
also suggests that a particularly rich medium is present at the location of
the western lobe.  High values of the $RM$ as found in IC~5063, are also measured
in Compact Steep Spectrum sources. Twenty percent of the B3-VLA CSS sample of
Fanti et al.\ (2004) have $RM \gta 1000$ rad/m$^2$. In these cases, the
Faraday screen most likely responsible for the $RM$ is the magnetised
interstellar medium. In most of these cases, the screen must
be confined to the nuclear environment. An exception to this seems to be the
high-redshift radio galaxy PKS B0529-549 (Broderick et al.\ 2006) where 
 $RM$ (--9600 rad/m$^2$) has been found. In this object, a gaseous
halo is likely responsible for the for the extreme $RM$.  The case of IC~5063
appears to be more similar to that of CSS sources. The presence of the high $RM$
seems to further support the idea that the NW radio lobe is actually hitting a
region of rich ISM.

Using the column density of the neutral hydrogen ($1 \times 10^{22}$ cm$^{-2}$
assuming \tspin\ = 1000 K) derived in Oosterloo et al.\ (2000), we can
estimate the mass of the \HI\ involved in the outflow.  Assuming that the
absorption covers the NW radio lobe ($\sim 1$ arcsec in size), we derive an
\HI\ mass of $3.6 \times 10^6$ $M_\odot$.  From the \HI\ data the mass outflow
rate can also be derived (see Morganti et al.\ 2005a).  Following Heckman
(2002) and Rupke et al.\ (2002), the mass outflow rate for IC~5063 is
estimated to be 35 $M_\odot$ yr$^{-1}$.  As discussed in that paper, this is a
substantial outflow rate that is comparable (albeit at the lower end of the
distribution) to that seen for starburst super-winds (Rupke et al.\ 2002)

\subsection{The ionised gas and the warm outflow}

The  optical spectra of IC~5063 confirm the high ionisation and the very
complex kinematics of the ionised gas (Colina et al.\ 1991, Wagner et
al.\ 1989).  The region where the complex gas kinematics is observed  is co-spatial with the
radio emission.  This  is very similar to what was previously found for other
Seyfert galaxies and suggests an effect of the radio plasma on the surrounding
medium, either via a direct interaction or via an expanding cocoon that
surrounds the radio jet (e.g. Mrk 3, Capetti et al.\ 1999).

\begin{figure}
\centerline{\psfig{file=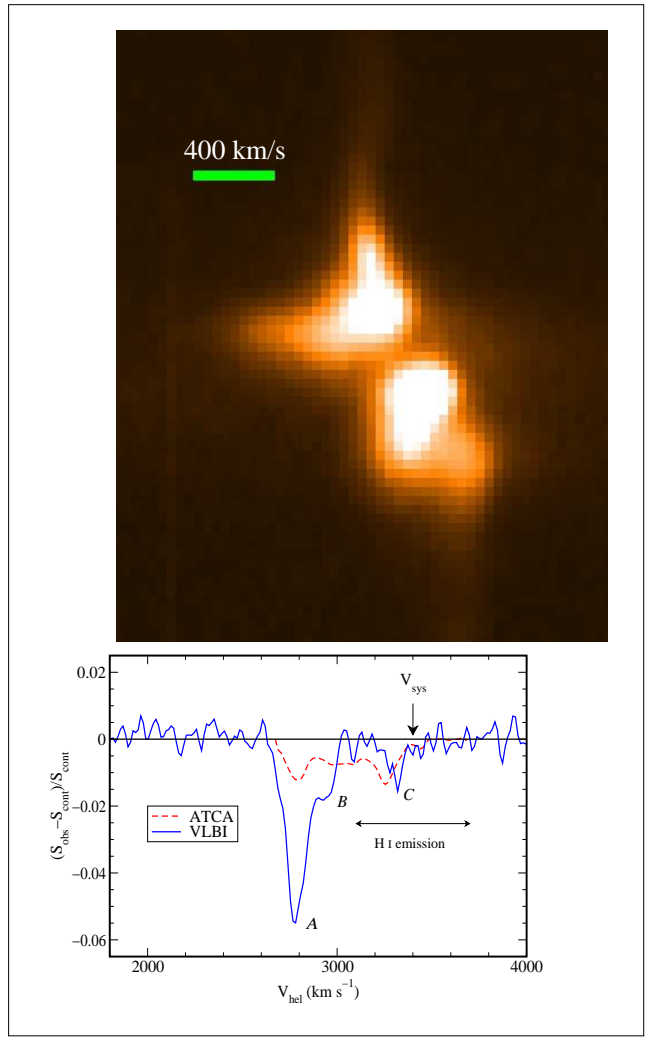,width=8.5cm,angle=0}}
\caption[]{Comparison between the width of the \HI\ absorption (white
  profile) and that of the ionised gas (from the {[\OIII]}5007\AA). The
  first order similarity between the amplitude of the blueshifted
  component is clearly seen.}
\label{fig:hi}
\end{figure}

Interesting in the context of gas outflows, is the presence of the highly
blueshifted emission lines in the region of the bright NW lobe. We interpret
this as a fast gas outflow occurring in this region.  Although we do not have
a similarly detailed image of the complete region for the \HI\ absorption, we
know from the VLBI data that the most blueshifted \HI\ absorption component is
also seen against the NW radio lobe.  The similarity between the kinematics of
this blueshifted wing of ionised gas and the \HI\ is illustrated in
Fig. {\ref{fig:hi}}.  Thus, a fast outflow is {\sl observed - at the same
location and with similar velocities - in both the atomic neutral and the
ionised gas}.  The fact that these outflows have such similar
characteristics indicates that the detection of outflows of neutral hydrogen
is likely not biased toward the strong radio continuum (i.e.\ one would see
the outflow occurring only where there is strong background
continuum). Nevertheless, we can not entirely rule out the idea that the
neutral outflow extends beyond the radio lobes and is driven by a different
mechanism (e.g. a starburst-induced super-wind).

Following Holt et al.\ (2006), we can estimate the mass outflow rate of the
warm, ionised gas and compare this with that of the neutral gas. We have
determined the \hb\ flux using a large aperture (1.5 arcsec) centred on the
outflowing region to include all flux in the outflowing component. This gives
a flux in the outflowing component of \hb\ of (2.34 $\pm$ 0.27) $\times$
10$^{-15}$ \dipso. For the reddening and density, we have used values
consistent with those presented in Table\ 3 for the outflowing region, namely
a reddening of $\ebv= 0.8$ and a gas density of $n_{\rm e}$ = 2000 cm$^{-3}$.
This gives a reddening corrected \hb\ flux of (3.53 $\pm$ 0.41) $\times$
10$^{-14}$ \dipso\ and the luminosity of the outflowing component in \hb\
(computed following the method of Holt et al.\ 2006) is L(\hb) = 9.77 $\times$
10$^{39}$ erg s$^{-1}$ cm$^{-2}$. A mass outflow rate of 0.08 \msun yr$^{-1}$
obtained. This is similar to the lower end of the range calculated for PKS
1549--79 (0.12-12 \msun yr$^{-1}$).  Thus, regardless the uncertainties in the
value of the density and as earlier found in other galaxies, the contribution
of the outflow of ionised gas is much lower than that of the \HI.  As
discussed in Morganti et al.\ (2005), the presence of fast {\sl neutral}
outflows indicates that after a strong jet-cloud interaction the gas can cool
very efficiently, as is indeed predicted by recent numerical simulations of
jets impacting on gas clouds (Mellema et al.\ 2002, Fragile et al.\ 2004,
Krause 2007). The relatively low outflow rate for the ionised gas suggests
that this cooling is indeed efficient and that most of the outflowing gas is
neutral.

 It is intriguing to see that, while neutral and ionised gas share, to first order, the same
location, only a relatively small part of the gas is ionised. Two
possibilities can be considered to explain this: \\
i) a large fraction of the jet-induced outflow is, for example, shielded from the AGN
ionizing continuum.  We could imagine a situation in which the
jet drills through the relatively dense ISM we know to exist in the galaxy,
staying relatively confined until it reaches a lower density part of the dust
lane structure where the lobes/jets can expand more freely, causing the
outflows we observe. Only a small fraction of the outflow is illuminated and
ionized by the AGN continuum through the hole drilled out by the initial
expansion of the radio jet through the denser inner ISM. In this case, high
resolution images of the ionized and neutral outflows would show that they
have different detailed spatial distributions, while still both associated
with the radio jets; \\
ii) It is also possible that, following cooling behind the
jet-induced shocks, the cooled cloud fragments are relatively large and
dense. In this case, the emission lines could be emitted by the outer skins of
the clouds photoionised by the AGN, with the masses/volumes of the skins much
smaller than the masses/volumes of the cloud fragments as a whole.
\\
Thus, in order to understand what is the situation we will need to image the exact
location of the full \HI\ outflow, something that at the moment is not available.

\subsection{Impact of the outflow}

It is important to examine whether the outflows have a significant impact
on the ISM of the host galaxy of IC~5063.  Given that the \HI\ outflow appears to
be the dominant one, in terms of mass outflow rate, we will perform the
calculations using the parameters found for the neutral hydrogen.  The kinetic
power associated with the neutral outflow is of the order of $7 \times
10^{42}$ erg s$^{-1}$.  This is derived by including the radial and the
turbulent component of the flow, the latter estimated from the width of the
line (see Holt et al.\ 2006 for more details).  To investigate whether the
characteristics of the observed outflows are interesting in the context of the
feedback model (Fabian 1999, di Matteo et al.\ 2005), we can use the values
derived from Nicastro, Martochia \& Matt (2003) for the black-hole mass and
accretion rate of IC~5063.  With a black-hole mass of $2.8 \times 10^8$
M$_\odot$, the Eddington luminosity of IC~5063 is $3.8 \times 10^{46}$ erg
s$^{-1}$, this means that the kinetic power of the outflow represents about
few $\times 10^{-4}$ of the available accretion power.  This result is
similar to that found for PKS~1549--79 (Holt et al.\ 2006).  However, unlike
PKS~1549--79, which accretes at a high Eddington rate, Nicastro et al.\ (2003)
found that IC~5063 accretes at a low rate ($\dot m \sim 0.02$, defined as the
ratio between the nuclear bolometric luminosity and the Eddington luminosity).
Thus, in IC~5063 the kinetic power of the outflow appears to be a relative
high fraction ($\sim 8$ \%) of the nuclear bolometric luminosity.  
Thus, the ratio between radiative luminosity ($L_{bol}$) and the kinetic
power in the neutral wind appears to be consistent with that required in the
Fabian (1999) model for the quasar feedback effect.  However,
the overall power in the outflow is much less than envisaged in the numerical
simulations of feedback in mergers, in which the quasar is accreting at close
to the Eddington rate (e.g. di Matteo et al. 2005).
Similar result is also found from the contribution of the  warm absorbers as observed in the narrow line
Syfert 1 NGC~4051 (Krongol et al. 2007).

\subsection{The SE side}

What is the situation on the other side (i.e.  on the SE side) of the nucleus?
Also there an intermediate-width component, with velocity width $\sim 600$
\kms\ is observed co-spatial with the radio emission.  This suggests that,
although not as extreme as on the NW side, also there the radio plasma
interacts with the surrounding medium.  As discussed above, the asymmetry seen
in the conditions of the medium surrounding the radio source may explain some
of the differences both in the radio morphology and in the kinematics of the
gas, as these asymmetries are likely to strongly affect the evolution of the
radio plasma.  We also see a component of ionised gas associated with the
regularly rotating large-scale gas disk.  In addition to all this, a second
component is seen, blueshifted compared to the quiescent gas.  It is hard to
explain this component only by a jet-cloud interaction and it is perhaps more
reminiscent of an expanding cocoon (see e.g.\ Mrk~3, Capetti et al.\ 1999). 
However, in the case of IC~5063, the second component is extending outside the
radio emission and this makes the interpretation more complicated. 
We note that the large-scale gas disk in IC~5063 is seen edge-on, so the
line-of-sight is intersecting the disk at various locations, resulting in a
large velocity widths and/or line splitting. This might explain the profiles
at the SE side.

\section{Conclusions}

We have presented new radio and optical observations of the central regions of
the Seyfert galaxy IC 5063. The radio data confirm that the fast outflow of
neutral hydrogen detected earlier, is indeed occurring off-nucleus, at about
0.5 kpc from the core. The optical spectra show that the kinematics of the
ionised gas deviates strongly from regular rotation in the region coincident
with the radio source. This supports the idea that the radio plasma ejected by
the Seyfert core is interacting with the ISM of IC 5063. Very broad,
blueshifted emissions lines are detected at the location of the NW radio lobe,
at the same location at which the neutral outflow is detected, pinpointing the
location of the strongest interaction. High rotation measures are detected in
the NW lobe, suggesting the presence of a high density medium near the NW
lobe.

Despite the strong kinematical evidence for an interaction between the radio
plasma and the ISM, we do not find, with the available limited data,
evidence for the dominant ionisation being due to this
interaction. Instead, photoionisation by the AGN is the most likely 
ionisation mechanism.

While the mass outflow rate of the neutral gas is substantial and is perhaps
large enough for AGN feedback effects to be important in IC 5063, the mass
outflow rate in the ionised gas is small compared to that seen in neutral
hydrogen. This can be explained by very efficient cooling of the outflowing
gas, as proposed in several models for jet-cloud interactions.

\begin{acknowledgements}

JH acknowledges a PPARC PhD studentship and a NOVA-Marie Curie Fellowship. She
also wishes to thank the Kapteyn Institute, Groningen, for its hospitality
where most of this work was done.

\end{acknowledgements}

\end{document}